\documentclass[a4paper]{article}

\usepackage{INTERSPEECH2019}
\usepackage{epstopdf}
\usepackage{cite}
\usepackage{url}
\usepackage{multirow}
\usepackage[table,xcdraw]{xcolor}
\usepackage{booktabs}
\usepackage{amsmath,amssymb,amsfonts}
\usepackage{graphicx}
\usepackage{textcomp}
\usepackage{xcolor}
\usepackage{times}
\usepackage{xcolor}
\usepackage{soul}
\usepackage[utf8]{inputenc}
\AtEndDocument{\par\leavevmode}
\usepackage{flushend}
\usepackage[section]{placeins}

\title{Identification of Dementia Using Audio Biomarkers}
\name{Rupayan Chakraborty, Meghna Pandharipande, Chitralekha Bhat, Sunil Kumar Kopparapu}
\address{ TCS Research and Innovation - Mumbai, INDIA}
\email{\texttt \{rupayan.chakraborty, meghna.pandharipande, bhat.chitralekha, sunilkumar.kopparapu\}@tcs.com}

\begin{document}

\maketitle
\begin{abstract}
Dementia is a syndrome, generally of a chronic nature characterized by a deterioration in cognitive function, especially in the geriatric population and is severe enough to impact their daily activities. Early diagnosis of dementia is essential to provide timely treatment to alleviate the effects and sometimes to slow the progression of dementia. Speech has been known to provide an indication of a person's cognitive state. The objective of this work is to use speech processing and machine learning techniques to automatically identify the stage of dementia such as mild cognitive impairment (MCI) or Alzheimer’s disease (AD). Non-linguistic acoustic parameters are used for this purpose, making this a language independent approach. We analyze the patients audio excerpts from a clinician-participant conversations taken from the Pitt corpus of DementiaBank 
database, to identify the speech parameters that best distinguish between MCI, AD and healthy (HC) speech. We analyze the contribution of various types of acoustic features such as spectral, temporal, cepstral their feature-level fusion and selection towards the identification of dementia stage. Additionally, we compare the performance of using feature-level fusion and score-level fusion. An accuracy of $82$\% is achieved using score-level fusion with an absolute improvement of $5$\% over feature-level fusion.
	
\end{abstract}

\section{Introduction}
\label{sec:intro}
Dementia is a syndrome in which the cognitive function of a person declines 
beyond what might be expected from normal ageing and is progressive in nature. 
It affects memory, thinking, orientation, comprehension, calculation, learning capacity, language, and judgement. 
Dementia is one of the major causes of disability and dependency among 
the geriatric population worldwide. The impact of dementia on the patients, carers and family can be overwhelming, affecting 
their physical, psychological, social and economic well-being. Timely diagnosis of dementia is imperative to provide in-time treatment.
Alzheimer's disease (AD) accounts for $60$-$80$\%  of all cases of dementia.
Manual diagnosis of AD requires specialized skills of neurologists and geriatricians through a series of 
cognitive tests such as the mini mental state examination (MMSE) \cite{FOLSTEIN1975189}. 
Other means of diagnosis involve collection and examination of cerebrospinal fluid from the brain and a magnetic
resonance brain imaging (MRI), that can be invasive and painful as well as expensive and tedious.
Hence a simple and non invasive approach is preferable. 
Longitudinal studies into aging demonstrates that speech is a good 
indicator of the cognitive function of a person \cite{DArcy2008SpeechAA}.
Speech has the advantage that it can be acquired non invasively in the form of a natural audio 
conversation with no additional stress on the person. 

Both linguistic (lexicon syntactic and semantic) and para-linguistic (acoustic) speech parameters
have been harnessed to estimate dementia status. 
Lexicosyntactic, acoustic, and semantic features were used to predict the MMSE scores
while stressing upon the need for longitudinal data collection for research purpose \cite{yancheva2015using}.
Earlier works used manual transcriptions to obtain linguistic features from dementia speech 
to classify into dementia stages \cite{fraser2016linguistic}, whereas 
recent works use automatic speech recognition to obtain the 
lexical and semantic features for this purpose \cite{2016ISMCI,sadeghian2017speech}. 
A combination of manual transcription based linguistic features and automatically extracted 
ASR based linguistic features have been used to detect dementia, wherein the 
automatically extracted features have shown to perform on par or better than manual transcription based features \cite{Weiner2017ManualAA}. 
N-gram based approaches have been used for automatic detection of AD from speech \cite{1626789,wankerl2017n}.
Working with acoustic features alone provides a 
language independent framework for dementia classification \cite{Satt2014SpeechbasedAA,al2016simple}.
A language independent system that uses a congnitive-task based framework along with nonverbal features 
to assess predementia is described in \cite{KONIG2015112}. 
An unsupervised system comprising voice activity detection (VAD) and speaker diarization
on conversational speech, followed by extraction of acoustic features and detection of early signs of  dementia 
demonstrates results that are comparable to a system using manual transcriptions \cite{Weiner2018}.
Word vector representations applied on spoken language to extract semantics 
from conversations that probe into people’s short and long-term memory have been shown to detect dementia \cite{Mirheidari2018}.
High accuracies have been reported for distinguishing between healthy control (HC) and AD speech \cite{al2016simple}. 
However, classification accuracy reduces when more than one stage of dementia is considered \cite{1626789}.

In this paper, we present a technique to identify dementia stages by classifying a given 
utterance into one of the three classes, namely, healthy (HC), with MCI or with AD using only
acoustic parameters from speech. We delve into the various types speech features such as spectral, temporal,
cepstral etc to identify the acoustic biomarkers that indicate the presence of dementia or AD. We explore early fusion of features and feature selection for identification of dementia stages. Moreover, we explore the late fusion
by using $2$ approaches,$(1)$ posterior classification probability fusion, $(2)$ Decision classifier. We also propose a method to balance the data to arrive at improved accuracies in identification of dementia stage. To the best of our knowledge classification of the Pitt corpus utterances into multiple stages of dementia has not been reported.

The rest of the paper is organized as follows. Section \ref{sec:method} describes the framework to identify dementia stage. In Section \ref{sec:scenario}, we describe the acoustic scenario and the database. Section \ref{sec:experiments} describes the 
experimental setup, results and analysis. We conclude in section \ref{sec:conclusion}.
 
\section{Framework for Dementia Identification}
\label{sec:method}
Dementia identification system consists of a feature extraction (FE) at the front end, followed by a conventional classifier to distinguish 
between HC, MCI and AD, given an audio utterance at the input. However, as found in literature \cite{Satt2014SpeechbasedAA}, ambiguities between classes (namely, HC-MCI and AD-MCI), motivated us to explore different audio biomarkers for dementia classification, individually as well as in various combinations. For combination, we follow two approaches, (1) Early fusion: audio biomarkers are fused at the front end before they are applied to the classifier, (2) Late fusion: by combining the posterior probabilities from each of the classifier that uses individual audio biomarkers for learning. Given a speech utterance $x(t)$, we extract audio feature vector $F_x$ by using a functional such as $\varphi (x)$; where 
$\varphi(.)$ can be a single or multiple functionals (both at low or high-level). For $n$ types of acoustic biomarkers, we represent the extracted features as $\{f_1, f_2,...f_n\}$.   
\begin{figure}
	\centering
	\includegraphics[width=0.4\textwidth]{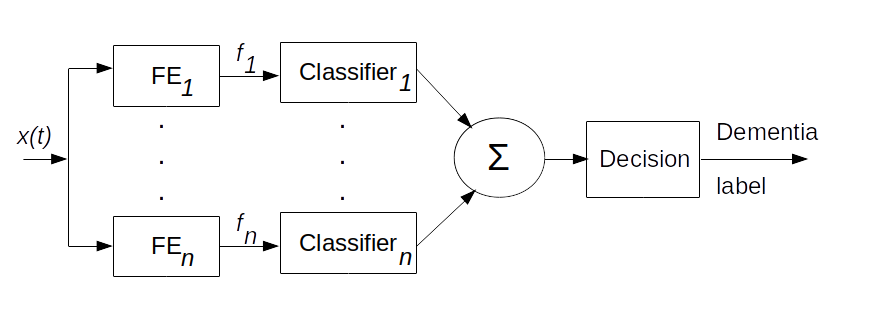}
	\caption{Dementia identification using early fusion of audio biomarkers}
	\label{figure:1}
\end{figure} 
\subsection{Early fusion}
In the early fusion approach, we concatenate the audio biomarkers (i.e. after the FE blocks) before they are applied to the classifier
(as shown in Figure \ref{figure:1}), which can be represented as,
\[
F = \{f_{1},f_{2},.....,f_{N}\}^{T}
\] 
Two methods of feature selection were adopted; one by selecting after concatenation, represented as    
\[
\Tilde{F} = \phi\{\{f_{1},f_{2},.....,f_{N}\}^{T}\}
\] 
and second by concatenating after selection of individual biomarkers, 
presented as,
\[
\Tilde{F} = \{ \phi (f_{1}),\phi (f_{2}),.....,\phi (f_{N}) \}^{T}
\]

\subsection{Late fusion}
\subsubsection{Probability fusion}
\begin{figure}
	\centering
	\includegraphics[width=0.45\textwidth]{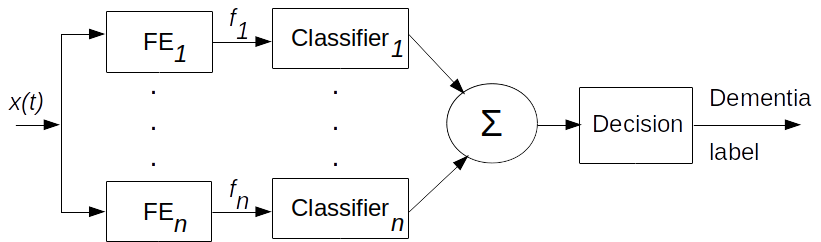}
	\caption{Dementia identification using late fusion}
	\label{figure:2}
\end{figure}

As depicted in Figure \ref{figure:2}, in late fusion, we extract several audio biomarkers, namely $f_1$, $f_2$, $f_n$, given an utterance $x(t)$. Then each classifier is trained by using each biomarker $f_i$ (or by using the selected features of the respective biomarker $\phi (f_i)$). Each classifier gives posterior probabilities $p(f_i|\varOmega^{j})$ as the output hypothesis, given $f_i$ feature vector as input. $\varOmega^{j}$ is the model created from the feature vector $f_i$ extracted from the training samples of dementia speech, $j$ is the index for dementia stages/classes to be considered. Then posterior probabilities from all the classifiers for each $\varOmega^{j}$ are added to obtain the class label based on the $\varOmega^{j}$ corresponding to the maximum probability. Decision process is defined as the following,
\begin{equation}
L= \operatorname*{arg\,max}_j \Big\{\sum\limits_{i=1}^{n}p(f_i|\varOmega^{j})\Big\}
\label{eq:1}
\end{equation}                             

\subsubsection{Decision fusion}
\begin{figure}[h]
	\centering
	\includegraphics[width=0.4\textwidth]{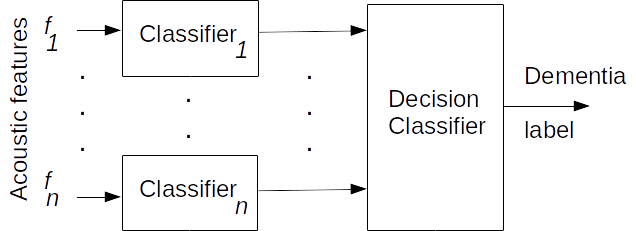}
	\caption{Decision classifier for dementia identification}
	\label{figure:3}
\end{figure}

Instead of using conventional probability fusion, we use a decision classifier to output the dementia class label. As shown in Figure \ref{figure:3}, the decision classifier takes probability scores which are hypothesized by the classifiers at the previous stage as input, and produces the class label at its output. The decision classification is represented as, 
\begin{equation}
L_D= \operatorname*{arg\,max}_j \Big\{ p(p_i|\varOmega^{j}_p)\Big\}
\end{equation}
where $\varOmega^{j}_p$ is the model created from all the output probability scores (hypothesized by all the classifier at the previous stage) for the $j^{th}$ class. 

\section{Acoustic scenario}
\label{sec:scenario}
\subsection{Database}
Pitt corpus \cite{Pitt} from the Dementia Bank dataset, collected at University of Pittsburgh School of Medicine
has been used. It comprises clinician-patient interviews in the form of audio, manual transcripts and subjective assessment
of the patient’s cognitive state as a longitudinal study over the span of four years. Data corresponds to four 
different linguistic-cognition tasks namely, picture description, fluency, recall and sentence construction. In this work we use the audio, transcription and subjective assessment from the picture description task, which is a verbal description of the
Boston Cookie Theft picture. It was recorded from people with different types of dementia with an age range from 49 to
90 years as well as from healthy (HC) subjects with an age range from 46 to 81 years. During the interviews, patients
were asked to discuss everything they could see happening in the cookie theft picture. 

We consider a sample data with a total of 597 recordings from 97 HC participants, 168 AD patients and 19 patients diagnosed 
with MCI. We use the speaker timing information provided in the transcripts, to remove the clinician turns from the 
recordings, retaining only the participant speech.

\begin{table}[ht]
\caption{Audio biomarkers for dementia classification}
	\label{table:features}
	\scalebox{0.8}{
	\begin{tabular}{|c|c|c|c|}
	    \hline
	    \textbf{Features} & \textbf{Dimensions} & \textbf{Details}&\textbf{Final} \\               
	    \hline \hline
	     $f_1$ & Cepstra(13)&(56 LLD + & \\
	     &Spectral(35)& $\delta$ + $\delta\delta$) & 6552\\             
	     &Energy(5)& * 39 functionals&\\
	     &Voice Probabilities(3)&&\\
	     \hline
	     $f_2$& Jitter - Shimmer& (3 LLD + $\delta$) & 114\\
	     &&* 19 functionals&\\
	     \hline
	     $f_3$& Speaking Rate&number of words per minute,& \\
	     &&number of syllables,& \\
	     &&speech duration, & 7\\
	     &&phonation time,&\\
	     &&number of pauses&\\
	     \hline
	     $f_4$&Posterior probabilities&probability score &\\
	     &&from the emotion acoustic & \\
	     &&models (anger,happy, & 7\\
	     &&neutral,sad,disgust,&\\
	     &&boredom,anxiety)&\\
	     
	     \hline
	\end{tabular}	
	}
\end{table}
  
\subsection{Audio biomarkers}
Speech utterances in the dataset are noisy, therefore we have used spectral subtraction to denoise it.
Details of audio biomarkers used for identification of dementia stages are provided in Table \ref{table:features}. 

$f_1$ feature set consists of cepstral, spectral, energy, voicing, their first ($\delta$) and second ($\delta \delta$) time derivatives as low-level descriptors (LLDs); 39 statistical functionals as high-level descriptors (HLDs). So the HLDs carry more relevant information than just using LLDs 
\cite{Geiger2013LargescaleAF}. Since the HLDs are statistics (up to fourth order) of LLDs over smaller frames (20 ms) in a spoken utterance, the dimension of the acoustic features remains the same (i.e. 6552) across utterances. We used {\em $emo\_ large$} configuration file from openSMILE toolkit for generating $f_1$ \cite{openSmile}.

$f_2$ feature set consists of jitter-shimmer features since they have traditionally been used in voice disorder classification \cite{TEIXEIRA20131112}. We consider $3$ pitch related LLDs, their $\delta$ and their $19$ statistical functionals as HLDs. These functionals do not apply to unvoiced regions, they are only applied to voiced regions. 
For jitter-shimmer features we use {\em INTERSPEECH 2010 Paralinguistic Challenge} configuration from the openSmile toolkit \cite{Schuller2016TheI2}.

Speaking rate related features are considered as $f_3$, which consist of number of words spoken per minute along with number of syllables, speech duration, phonation time, number of pauses are also calculated for each utterance\cite{speakrate}. Based on our observation of the patient audio and the typical known characteristics of dementia, speaking rate helps in distinguishing between the healthy and dementia speech \cite{phdthesis}. 

We consider $f_4$ as a feature set, that comprises of posterior probabilities, which we get using pre-trained emotion models. It has been observed that individuals with dementia express extreme emotional distress indicating the importance of anxiety and apathy in dementia, that can be assessed using speech parameters \cite{doi:10.1002/gps.4870}. The posterior probabilities are hypothesized by a emotion classifier which is trained by using utterances from the emotional speech database, EmoDB \cite{EmoDB}. The probability scores are expected to capture the variation of the affective contents, given a speech utterance as input. This classifier caters to 7 different emotions namely, happy, angry, sad, neutral, disgust, boredom and anxiety. The probability scores corresponding to the emotions are used as a feature vector, giving rise to a feature vector of length 7.
\section{Experiments}
\label{sec:experiments}
\subsection{Experimental setup}

The objective of this work is to analyze the implications of various types of acoustic biomarkers in the identification of dementia and distinguishing between dementia stages. As a first step, we apply individual speech feature sets described in Table \ref{table:features} on a Random Forest classifier with 100 trees along with 10-fold cross validation. 
Further, feature selection is carried out using the attribute evaluator CfsSubsetEval with the search method 
GreedyStepwise specified in the Weka toolkit \cite{DBLP:soft/weka}.
Two methods of early fusion in the feature space are carried out as described in Section \ref{sec:method}.
Finally, the best performing combination of features was applied to $5$ different types of classifiers namely Random Forest, Multilayer Perceptron (MLP), Sequential Minimal Optimization (SMO) based Support Vector Classifier, 
SimpleLogistic and BayesNet provided in the Weka toolkit. 

We also explored the late fusion or score fusion wherein we used the posterior probability scores 
of a classifier to arrive at the class decision as described in Section \ref{sec:method}.
Consistent with the difficulty in identifying early stages of dementia, the data in Pitt corpus comprises fewer participants with MCI as compared to HC and AD participants. This gives rise to an unbalanced data set with higher number of utterances for AD and HC and around half the amount for MCI. In order to eliminate the bias of data from training the classifiers, we balance the data using Weka preprocessing filter \textit{SpreadSubsample}, with instance weights adjusted to maintain total weight per class which produce random subsample of a dataset based on number of instances.
\begin{table}[t]
	\caption{Performance of individual feature sets $f_1$, $f_2$, $f_3$, $f_4$}
	\label{table:featureLevel}
	\scalebox{0.9}{
		\begin{tabular}{|c|c|c|c|c|}
			\hline
			{Features}&Dimension	&Pr	&Re	&F-Score\\
			\hline
			$f_1$-Cepstral	&1521		&56.9	&56	&55\\	
			$f_1$-Spectral	&4095		&68.5	&66.5	&66.43\\
			$f_1$-Energy	&585		&57.8	&55.5	&54.2\\
			$f_1$-VP		&351		&57.1	&57.7	&56.4\\
			\hline
			$f_1$		&6552		&70.1	&68.5	&\textbf{65.9}\\
			$f_2$		&114		&62.3	&62	&61\\
			$f_3$		&7		&52.1	&50.5	&49.81\\
			$f_4$		&7		&45.5	&49.2	&46.5\\
			\hline
		\end{tabular}
	}
\end{table}
\subsection{Results and Discussion}
\label{sec:results}
Performance of a 3-class Random Forest classifier for each individual feature set is as shown in Table \ref{table:featureLevel}.
It can be seen that the F-score for the feature set $f_1$ and $f_2$ are well above that for $f_3$ and $f_4$. 
The best score was obtained for the feature set $f_1$ as 66\% F-score.
This indicates that although a person's cognitive state does have a bearing on speaking rate and emotion in speech, 
it is not sufficient for classification on it's own. Therefore, we retain these features and use them in combination with 
the feature sets $f_1$ and $f_2$ through the early fusion process described in \ref{sec:method}. 
Table \ref{table:earlyFusion1} shows the performance of early fusion of feature sets along with feature selection process 
applied. The best overall F-score is achieved when feature selection is done on individual sets of features and then 
concatenated. This is also the best performing set up for each individual class with the overall F-score improving by 10\%.
While the classification for AD and HC remain high, the performance for MCI improves significantly when 
early fusion of features is employed. It is this improvement that contributes to the overall improvement of the system,
It is also crucial to improve the F-score for MCI classification from clinical perspective in order to provide adequate 
attention to patients at this stage to help slow the disease progress. 
\begin{table}[h]
	\caption{Early Fusion - Random Forest Classifier}
	\label{table:earlyFusion1}
	\scalebox{0.77}{
		\begin{tabular}{|c|c|c|c|c|c|}
			\hline
			Features		&Dim	&Class	&Pr	&Re	&F-Score\\
			\hline
			&	&AD	&70.4	&77.9	&73.9\\
			&       &MCI	&64.6	&25.2	&37.4\\		
			\{$f_1$,$f_2$,$f_3$\}	&6673	&HC	&72.5	&78.0	&70.7\\
			\cline{3-6}
			&	&Overall&67.8	&67.7	&65.6\\
			\hline
			&	&AD	&77.8	&81.8	&79.8\\
			&	&MCI	&73.3	&40.9	&53.4\\
			$\phi$\{{$f_1$,$f_2$,$f_3$}\}	&54	&HC	&77	&86.2	&79.2\\
			\cline{3-6}
			&	&Overall&75.9	&75.7	&74.5\\
			\hline
			&	&AD	&67.9	&77.9	&72.6\\
			&	&MCI	&66.2	&22.6	&34\\
			\{$f_1$,$f_2$,$f_3$,$f_4$\}&6680	&HC	&68.4	&77.6	&71.4\\
			\cline{3-6}
			&	&Overall&67.3	&67.2	&64.7\\
			\hline
			&	&AD	&79.6	&83.4	&81.5\\
			&       &MCI	&71.9	&39.1	&52.3\\
			$\phi${\{$f_1$,$f_2$,$f_3$,$f_4$\}}&60	&HC	&78.9	&86.2	&78.4\\
			\cline{3-6}
			&	&Overall&76.5	&76	&74.7\\
			\hline
			&	&AD	&79.8	&83	&81.4\\
			&	&MCI	&73.3	&43.5	&57.1\\	
			\{$\phi$\{$f_1$\},$\phi$\{$f_2$\},$\phi$\{$f_3$\},$\phi$\{$f_4$\}\}	&86	&HC	&83.3	&87.5	&79.8\\
			\cline{3-6}
			&	&Overall	&78	&77.2	&\textbf{76.1}\\
			\hline
		\end{tabular}
	}
\end{table}

\begin{table}[h]
	\caption{Multiple classifiers for the Early fusion configuration }
	\label{table:earlyFusion2}
	\scalebox{0.85}{
		\begin{tabular}{|c|c|c|c|c|}
			\hline
			{Classifiers}		&Class	&Pr	&Re	&F-Score\\
			\hline
			&AD	&81.3	&82.6	&82\\
			&MCI	&57.5	&53	&55.2\\
			MLP		&HC	&78.5	&80.2	&79.3\\
			\cline{2-5}
			&Overall&75.7	&76	&75.8\\
			\hline
			&AD	&77.3	&83.4	&80.2\\
			&MCI	&54.4	&32.2	&40.4\\
			SMO		&HC	&76.1	&84.9	&80.2\\
			\cline{2-5}
			&Overall&72.4	&72.2	&72.6\\
			\hline
			&AD	&77.9	&85	&81.3\\
			&MCI	&58.2	&40	&47.7\\
			SimpleLogistic	&HC	&76.7	&81	&78.8\\
			\cline{2-5}
			&Overall&73.7	&74.8	&73.8\\
			\hline
			&AD	&81.9	&80.2	&81\\
			&MCI	&70.6	&62.6	&66.4\\
			BayesNet		&HC	&76	&81.9	&78.8\\
			\cline{2-5}
			&Overall&77.4	&77.5	&77.4\\
			\hline
		\end{tabular}
	}
\end{table}
We assess the robustness of the combination of features selected that gives the best performance (\{$\phi$\{$f_1$\},$\phi$\{$f_2$\},$\phi$\{$f_3$\},$\phi$\{$f_4$\}\}), by applying these features to various classifier configurations as shown
in Table \ref{table:earlyFusion2}. While the performance of these classifiers remained roughly the same for AD and HC,
classification of MCI showed some variation, with \textit{BayesNet} giving the best F-score at 66\%.

Further, we used the posterior probabilities from individual classifiers trained on selected features of each of the 
feature sets $f_1$, $f_2$, $f_3$ and $f_4$ to arrive at a class decision. We adopted two methods for this purpose as described in Section \ref{sec:method}.
Table \ref{table:LateFusion1} shows the performance for both types of late fusion methods when using a Random Forest classifier. Additionally, we used 4 other types of classifiers as decision classifiers in the late fusion method. 
Performance for these classifiers when we consider unbalanced and balanced is as shown in Table \ref{table:LateFusion2}.
This clearly shows that for each type of classifier, the performance improves when the data is balanced across classes.
Best overall performance was obtained for the SMO classifier at 82\%.

\begin{table}[h]
        \caption{Late-Fusion for Random Forest Classifier}
	\label{table:LateFusion1}
	\scalebox{0.9}{
	\begin{tabular}{|c|c|c|c|c|}
			\hline
			\textbf{Method}&\textbf{Class}&\textbf{Pr} & \textbf{Re} & \textbf{F-Score}\\ 
			\hline
			&AD&80.3&82.2&81.3\\
			&MCI&67.4&53.9&59.9\\
			$\operatorname*{arg\,max}_j \Big\{\sum\limits_{i=1}^{n}p(f_i|\varOmega^{j})\Big\}$&HC&74.7&80.2&77.3\\
			\cline{2-5}
			&Overall&75.7&76&75.6\\
			\hline
			&AD&82.7&83.4&83.1\\
			&MCI&67.3&59.1&63\\
			$\operatorname*{arg\,max}_j \Big\{ p(p_i|\varOmega^{j}_p)\Big\}$&HC&75.4&79.3&77.3\\
			\cline{2-5}
			&Overall&77.0&77.2&77\\
		\hline	
		\end{tabular}
		}
\end{table}

\begin{table}[ht!]
        \caption{Dementia identification using Late fusion method}
	\label{table:LateFusion2}
	\scalebox{0.75}{
	  \begin{tabular}{|c|c|c|c|c||c|c|c|}
	  \hline
	      \multirow{2}{*}{Classifier}&{Class}& \multicolumn{3}{c||}{Unbalanced} &\multicolumn{3}{c|}{Balanced}\\
\cline{3-8}
&&Pr&Re&F-Score&Pr&Re&F-Score\\
\hline
&AD&83.8&83.8&83.8&85.6&82.6&84.1\\
&MCI&68&59.1&63.3&64.1&67.8&65.9\\
MLP&HC&75.7&80.6&78.1&77.5&78.3&77.9\\
\cline{2-8}
&Overall&77.6&77.8&77.6&78.4&78.1&78.2\\
\hline
&AD&84.8&83.8&84.3&86.9&87&86.9\\
&MCI&80.4&64.3&71.5&79.6&63.5&70.6\\
SMO&HC&77.1&85.5&81.2&78.3&86.1&82\\
\cline{2-8}
&Overall&81&80.8&80.7&82.2&82.1&\textbf{81.9}\\
\hline
&AD&81.2&83.8&82.5&84.7&87.8&86.2\\
&MCI&75.8&62.6&68.6&74.2&67&70.4\\
SimpleLogistic&HC&76.6&80.6&78.6&80.2&80.9&80.6\\
\cline{2-8}
&Overall&78.4&78.5&78.3&81&81.1&81\\
\hline
&AD&84&80.6&82.3&84.8&81.7&83.2\\
&MCI&61.9&67.8&64.7&63.8&69.6&66.5\\
BayesNet&HC&76.2&75.9&76&76.1&75.7&75.9\\
\cline{2-8}
&Overall&76.7&76.3&76.5	&77.4&77.1&77.2\\
\hline
\end{tabular}
}
\end{table}	

\section{Conclusion}
\label{sec:conclusion}
Early diagnosis of dementia is essential to provide timely treatment in order to
alleviate the effects and sometimes to slow the progression of dementia. Speech has 
been known to provide an indication of a person's cognitive state. Through this work we explore 
the usefulness of acoustic biomarkers and demonstrate a mechanism to identify dementia stages effectively using these
biomarkers. Traditional 3-class classifier with no feature selection process, achieved an overall F-score of 66\%, 
whereas by using early fusion of features, this improves to 77\%. Across the experiments with various audio biomarkers and classifiers, it was observed that the classification for MCI performed poorly as compared to the AD and HC category.
This is consistent with the fact that distinguishing early stage dementia from healthy speech is challenging even for trained experts. This leads to a biased dataset with fewer samples for MCI speech. We addressed this bias in data distribution by balancing the data across classes. Using late fusion mechanism on balanced data, we were able to further improve the dementia stage identification to 82\% which is a significant improvement using only acoustic biomarkers.
\bibliographystyle{IEEEtran}
\bibliography{dementiaClassification}

\end{document}